\documentclass[%
 reprint,
 amsmath,amssymb,
 aps,
]{revtex4-1}

\usepackage{graphicx}% Include figure files
\usepackage{dcolumn}% Align table columns on decimal point
\usepackage{bm}% bold math
\usepackage{natbib}
\usepackage{float}

\begin{document}

\preprint{APS/123-QED}

\def\dbar{{\mathchar'26\mkern-12mu d}}
\newcommand{\angstrom}{\mbox{\normalfont\AA }}

\title{Thermodynamics of a Simple Three-Dimensional DNA Hairpin Model}
\author{Kellan Kremer}
\author{Kyle Jensen}
\author{Erin Boggess}
\author{Walker Mask}
\author{Tony Saucedo}
\author{JJ Hansen}
\author{Ian Appelgate}
\author{Taylor Jurgensen}
\author{Aaron Santos}
 \email{aaron.santos@simpson.edu}
\affiliation{Chemistry and Physics Department, Simpson College \\
701 North C Street, Indianola, IA 50125
}%

\date{\today}

\begin{abstract}
We characterize the equation of state for a simple three-dimensional DNA hairpin model using a Metropolis Monte Carlo algorithm.
This algorithm was run at constant temperature and fixed separation between the terminal ends of the strand.
From the equation of state, we compute the compressibility, thermal expansion coefficient, and specific heat along with adiabatic path.
\begin{description}
\item[PACS numbers]
\end{description}
\end{abstract}

\pacs{Valid PACS appear here}
\maketitle

%%%%%%%%%%%%%%%%%%%%%%%%%%%%%%%%%%%%%%%%%%%%%%%%%%%%%%%%%
\section{Introduction}
%%%%%%%%%%%%%%%%%%%%%%%%%%%%%%%%%%%%%%%%%%%%%%%%%%%%%%%%%

\begin{figure}[h]
\includegraphics[scale=.28]{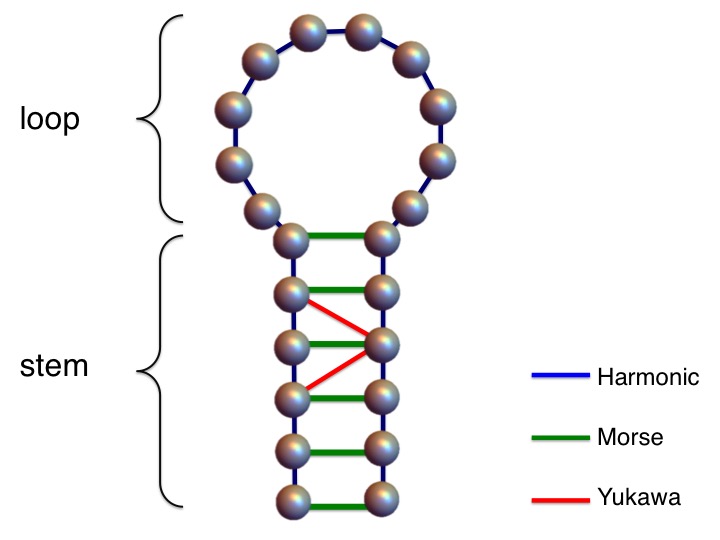}
\caption{\label{Fig1} A DNA hairpin showing the stem and loop. Harmonic, Morse and Yukawa bonds are shown in blue, green, and red, respectively. For ease of viewing, only two Yukawa bonds are shown.}
\end{figure}

DNA is one of many self-assembling biological polymers~\cite{winfree_design_1998}.
These molecules form different arrangements depending on their sequence~\cite{rothemund_folding_2006}.
For DNA, one of these arrangements is the hairpin-loop, a secondary structure formed when single stranded DNA (ssDNA) has two complimentary sequences that fold on top of each other in the shape of a hairpin~\cite{bonnet_kinetics_1998}.
The loop of the hairpin is comprised of single stranded DNA, while the stem is formed by double stranded DNA connected by hydrogen bonds between complementary base pairs, as illustrated in Fig.~\ref{Fig1}.
The hairpin structure is not static, but fluctuates predominately between two unique conformations: the open state and the closed state \cite{bonnet_kinetics_1998}.  
The DNA predominantly exists in the closed state at temperatures below its melting temperature, allowing the hydrogen bonds in the stem to form.  
As the temperature increases above the melting temperature, the stem denatures and the structure behaves like linear polymer.  

Since DNA has only four different monomers in its primary sequence, there is a relatively high probability that complimentary sequences within the same strand will be located close enough to bind to each other. 
As such, hairpin structures are fairly common.
For instance, hairpins are typically formed during replication~\cite{chen_hairpins_1995,voineagu_replication_2008}.  	
Furthermore, DNA hairpins play a multitude of biological roles, including the regulation of gene expression 
\cite{roth_v_1992, gottesfeld_regulation_1997, zazopoulos_dna_1997, smith_gene_2000, mccaffrey_gene_2002, yu_rna_2002}, 
DNA recombination \cite{froelich-ammon_site-specific_1994, kennedy_tn10_1998, ma_hairpin_2002, lengsfeld_sae2_2007}, 
and mutagenesis \cite{collins_instability_1981, trinh_influence_1993, wang_non-b_2006, wells_non-b_2007, kruisselbrink_mutagenic_2008}.
In addition to its biological importance, DNA has many applications in nanotechnology, including 
nanomedicine~\cite{yan_tang_self-stabilized_1993, kagan_nanomedicine_2005, liu_nanomedicine_2007, roy_optical_2005, maojo_nanoinformatics_2010, chhabra_dna_2010, campolongo_dna_2010, lopez_catalytic_2010, ma-ham_apoferritin-based_2011}
and nanorobotics~\cite{hamdi_dna_2008, reif_autonomous_2009, elbaz_dna_2012, douglas_logic-gated_2012, jani_dna_2013, popov_avoidance_2014}. 
DNA hairpins in particular are used to make 
biosensors~\cite{tyagi_molecular_1996, mao_studies_2003, du_hybridization-based_2003, jin_hairpin_2007, zhang_electrogenerated_2008-1}, 
DNA computers~\cite{adleman_molecular_1994, liu_dna_1998},
and shape shifting smart materials \cite{goodman_reconfigurable_2008}.

To better understand the thermodynamic and statistical properties of both engineered and biological DNA-based structures, researchers have employed a variety of single molecules force measuring techniques. These include 
nanopores~\cite{storm_fast_2005, van_dorp_origin_2009, keyser_direct_2006, keyser_optical_2006}, 
optical tweezers~\cite{keyser_direct_2006, keyser_optical_2006, wang_stretching_1997, bockelmann_unzipping_2002, tropini_multi-nanopore_2007,neuman_single-molecule_2008,bustamante_single-molecule_2000},
and AFM~\cite{lee_direct_1994, boland_direct_1995, strunz_dynamic_1999, rief_sequence-dependent_1999, bustamante_single-molecule_2000, rouzina_force-induced_2001}.
Force measurement experiments typically involve anchoring one portion of the molecule while another portion is pulled with a force that can be either measured or derived.

In addition to single molecule experiments, simple coarse grained models can elucidate many qualitative features about the behavior of DNA at the nanoscale.
DNA modeling originated with simple Ising-like models in the 1960s~\cite{richard_polandscheraga_2004, fisher_walks_1984}. 
These simplified, two-state models only characterize a base pair as existing in an open or closed state.
As such, these models excel at predicting the thermodynamic behavior of systems containing a large number of base pairs \cite{peyrard_modelling_2008}, however, they fail when fine resolution is required to detect large-amplitude fluctuations at temperatures below that of denaturation \cite{dauxois_entropy-driven_1993}.    
Coarse grained models provide a better avenue for understanding physics at the mesoscopic level~\cite{knotts_iv_coarse_2007}. 
Computational methods for analyzing these models are diverse, including Monte Carlo methods, Lattice Boltzmann methods, Brownian dynamics, and molecular dynamics \cite{kenward_brownian_2009,frederickx_anomalous_2014}.   
One model, the Peyrard Bishop (PB) model, simplifies the structure of DNA by representing hydrogen bonds and stacking interactions as Morse and Harmonic bonds, respectively~\cite{peyrard_statistical_1989}. 
Cuesta-L\'{o}pez, Peyrard, and Graham (CLPG) further extended the PB Model to account for the two-dimensional position of each base, rather than simply the separation between bases \cite{cuesta-lopez_model_2005}.
This modification permitted the formation hairpin structures, while still maintaining a great deal of simplicity.
Using this model, CLPG analyzed the melting characteristics of hairpins under various circumstances to examine the role strand rigidity and other properties play in melting. 

While CLPG focused primarily on melting, the simplicity of their model makes computing other thermodynamics properties via simulation straightforward, as many states can be sampled in a relatively short computational run time.
Moreover, while the model may not be detailed enough to quantitatively capture fine details of a DNA hairpin, it's simplicity may make it generally applicable to qualitatively describing the features of many types of folded polymers.
In what follows, we examine the thermodynamic properties and response functions for the CLPG model using data obtained from Metropolis Monte Carlo simulations.

This paper is organized as follows. 
In section 2, we describe a three dimensional version of the CLPG hairpin model that we simulated as well as the Monte Carlo algorithm we used to determine the thermodynamic properties of the model. 
In section 3, we discuss the results including fitted functional forms for approximate equations of state along with notable features of the thermodynamic response functions.
In section 4, we conclude by discussing possible implications and applications of the work described.

%%%%%%%%%%%%%%%%%%%%%%%%%%%%%%%%%%%%%%%%%%%%%%%%%%%%%%%%%
\section{Methods}
%%%%%%%%%%%%%%%%%%%%%%%%%%%%%%%%%%%%%%%%%%%%%%%%%%%%%%%%%

We simulated a three-dimensional version of the DNA hairpin model originally introduced by CLPG~\cite{cuesta-lopez_model_2005}.
While the CLPG model was originally simulated in two dimensions, the forms for the potentials make it straightforward to extend the model to three dimensions. 
In this model, each of the $N=24$ beads represents a nucleotide comprised of a base, sugar, and a phosphate group.
The 6 beads at each end of the strand were assigned to the stem and allowed to bond to the complimentary bead at the opposite end.
This left 12 beads in the loop of the hairpin.
We chose to keep all parameters in the three dimensional model within an order of magnitude as the original CLPG model, as this ensured a realistic melting temperature.
Below we describe the model's energetic potentials and define the parameters.

\subsection{Model}%%%%%%%%%%%%%%%%%
The CLPG model describes stacking interactions and hydrogen bonding between complimentary bases in the stem via a Morse potential.
This potential takes the form
\begin{align}
V_M=D\left[e^{-\alpha\left(\left|\vec{r}_i-\vec{r}_j \right|-d_0  \right)}-1\right]^2,
\end{align}
where $d_0=10$ \r{A}  is the equilibrium separation between paired bases in the stem, $D=0.22$ eV is the well depth, $\alpha=4.45$ \r{A}$^{-1}$ is the inverse well width, and $\vec{r}_i$ and $\vec{r}_j$ are three dimensional position vectors to bases $i$ and $j$, respectively.
Each base pair in the stem of the hairpin interacts through a Morse potential.

The energy between adjacent beads in the strand is described by a harmonic potential of the form
\begin{align}
V_H= K_s \left(|\vec{r}_i-\vec{r}_j |-r_0 \right)^2,
\end{align}
where $K_s=0.22$ eV/\r{A}$^2$ represents the stiffness of the spring and $r_0=6.0$ \r{A} represents the average interparticle distance.  

It is known that the flexibility of a DNA strand is dependent on its sequence~\cite{goddard_sequence_2000}.
To account for the role sequence-dependent base stacking plays in the flexibility of the strand, CLPG introduced a rigidity potential,
\begin{align}
V_R=K_r (1+\cos\theta_i ),
\end{align}
where $K_r$ is the strength of the rigidity potential and $\theta_i$ is the angle formed by bases $i-1$, $i$, and $i+1$.  
This potential influences the shape of the loop, with larger rigidity strengths $K_r$ creating more rounded loops and small $K_r$ resulting in more twisted loops.
Here, we have chosen to examine systems with a fairly weak rigidity strength, $K_r=0.05$ eV.

Finally, CLPG included a Yukawa potential to eliminate shear distortion within the hairpin stem.  
This potential, which can be written as
\begin{align}
V_{Yuk}=k_{yuk} \left(\frac{e^{-C_{yuk} |r_i-r_j | }}{|\vec{r}_i-\vec{r}_j | }\right),
\end{align}
provides a repulsive force between base $i$ and any base $j$ adjacent to its complimentary pair on the opposite side of the stem.    
Here, $k_{yuk}=50.0$ eV$\cdot$\r{A} represents the strength of the Yukawa potential and $C_{yuk}=0.40$ \r{A}$^{-1}$ represents the inverse Debye screening length. 

For brevity, we will henceforth drop units on numerical values listed in the text.
Toy models like the CLPG model used here are only expected to produce qualitatively accurate results. 
As such, exact numerical values are not expected to agree quantitatively with experiments, and the inclusion of units is, perhaps, misleading.
Readers who wish to compare the quantitative results to experimental values are free to use dimensional analysis taking the units eV and \r{A} as a base.

\subsection{Simulation Method}%%%%%%%%%%%%%%%%%
To simulate the statistics of the DNA strand we use a Metropolis Monte Carlo (MC) algorithm.
Applied to our model, each time step of the algorithm proceeds as follows:
\begin{enumerate} 
\item Generate a trial move by choosing a random bead and displacing it in a random direction by a random distance between 0 and $\sqrt{3}$.  
\item Compute the change in energy $\Delta E$. 
\item Accept the move with a probability given by the Boltzmann distribution,
\begin{align}
P_{acc}=e^{-\Delta E/k_BT}
\end{align}
where $T$ is the temperature and $k_B$ is Boltzmann's constant.
\end{enumerate}

In each run, the terminal bases were separated by a distance $x$, which was held fixed throughout the simulation.
By holding the position fixed, we mimic the situation where an AFM tip or some other positioning device constrains the ends of the polymer.
Thermodynamically, this corresponds to the {\it N-x-T} ensemble.

\begin{figure}[t]
\includegraphics[scale=.35]{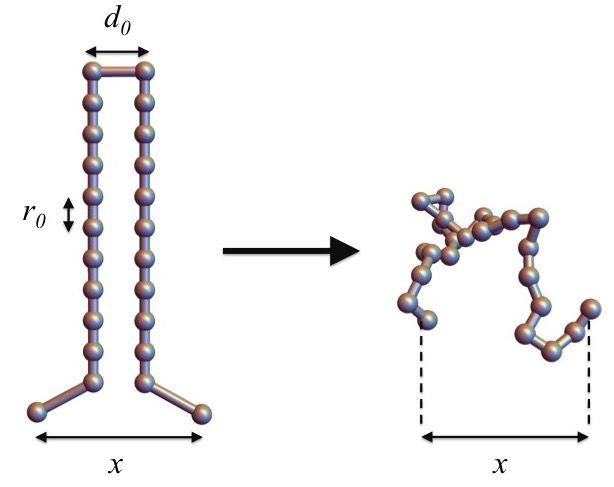}
\caption{\label{Fig2} The starting configuration of a DNA hairpin (left) and the state of the hairpin after equilibration (right). The terminal bases are fixed with a separation distance $x$ throughout the simulation.}
\end{figure}

We began each run by initializing the DNA in a ``bent'' hairpin state as illustrated in Fig.~\ref{Fig2}. 
In this configuration, the strand was bent at its center to produce two segments of equal length, each containing 12 beads. 
With the exception of the terminal bases, adjacent bases within a segment were separated by a distance $r_0$,  while the two segments themselves were arranged in parallel lines separated by a distance $d_0$.
This arrangement ensured that bases within each segment would have zero contribution to the energy from the harmonic and rigidity potentials, while bases in the stem would be bonded at the minimum of the Morse well.
Bases at the kink in the hairpin would exhibit a large energy contribution arising from the rigidity potential, but this high energy conformation generally did not survive an equilibration period before recording data.

We initialized each MC run with a different random number seed to ensure the sequence of MC moves was statistically independent from that of other runs. 
Since a 24 particle system size is fairly small, we were able to run the simulations for a total of $10^8$ MC steps, of which the first $2.5\times10^7$ were spent allowing the system to equilibrate.
After the equilibration period, we recorded the state of the system every $5\times10^5$ MC steps. 
From these states, we then computed the energy and the net force on the first bead.
It is only necessary to compute the force along the separation vector in the $x$ direction, since the $y$ and $z$ components of the force will average to zero by symmetry.
After computing the energy and force for each of the 175  states recorded throughout the run, we then computed the average energy and average force on the terminal bases for the run.
 
To see how the system behaves over a wide range of conformations, we ran simulations with separation distances ranging from $x=9.5$ to $x=11.9$ at intervals of 0.3, and again from  $x=12$ to $x=152$ at intervals of 20.
These values ensured there was sufficient data to determine a decent fit for the state equations both in the Morse well where the energy and force change rapidly, and outside the Morse well where these functions change slowly.
For each one of these separation distances, we ran simulations at temperatures ranging from 200-400 at intervals of 4.

%%%%%%%%%%%%%%%%%%%%%%%%%%%%%%%%%%%%%%%%%%%%%%%%%%%%%%%%%
\section{Results}
%%%%%%%%%%%%%%%%%%%%%%%%%%%%%%%%%%%%%%%%%%%%%%%%%%%%%%%%%

\subsection{Equations of State}%%%%%%%%%%%%%%%%%%%%%%%%%%%%%%%%%%%%%%%%%%

\begin{figure}[b]
\includegraphics[scale=0.35]{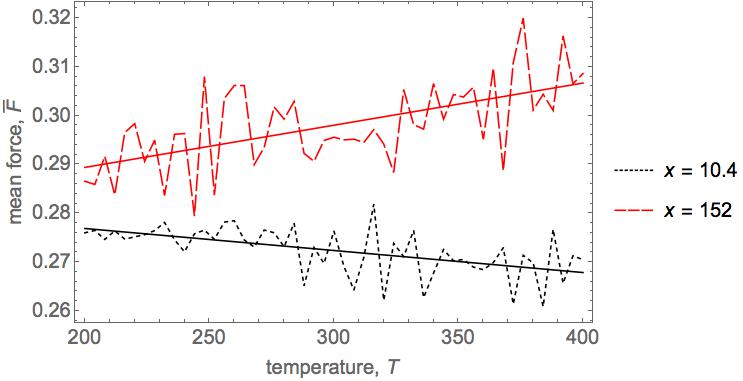}
\caption{\label{fig:FbarVsTemp10-152} A plot of the mean force $\bar{F}$ acting on the terminal bases of the hairpin versus temperature  $T$ for separation values of $x = 10.4$ (red, dashed) and $x=152$ (black, dotted). The straight lines represent linear fits to the data.}
\end{figure}

\begin{figure}[H]
\includegraphics[scale=0.35]{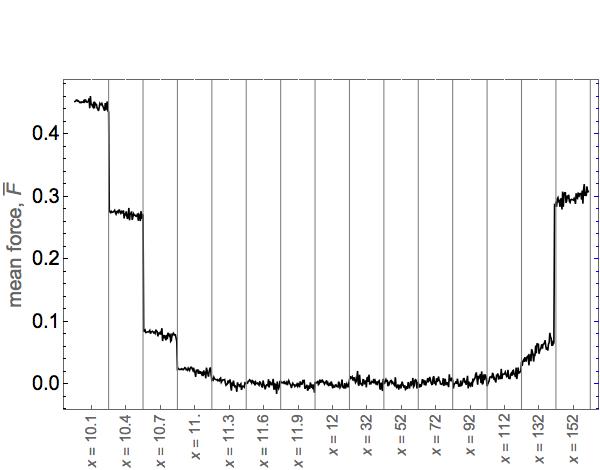}
\caption{\label{fig:FbarVsTempAll} A plot of the mean force $\bar{F}$ acting on the terminal bases of the hairpin versus temperature  $T$ for terminal base pair separations ranging from $x = 10.1$ to $x=132$. }
\end{figure}

In Fig.~\ref{fig:FbarVsTemp10-152}, we plot the mean force $\bar{F}$ on the terminal bases as a function of temperature $T$ for terminal base pair separation values $x = 10.4$ (red, dashed) and 152 (black, dotted). 
As can be seen from the linear fits to the data, the slope changes significantly from negative to positive between these two separation values.
This suggests the slope has an appreciable dependence on the separation between the terminal bases.
A trend in the slope can be observed by looking at $\bar{F}$ versus $T$ plots for many separation values simultaneously. 
In Fig.~\ref{fig:FbarVsTempAll}, we plot $\bar{F}$ versus $T$ for multiple values of separation $x$. 

In Fig.~\ref{fig:FbarVsTempAll}, one  observes negatively sloping curves at small separations.
This slope steadily increases to positive values as $x$ increases.
Moreover, there is a significantly larger mean force for $x\leq11$ and $x\geq132$.
This can be easily understood by noting that the force is largest for small separations, i.e. when the terminal bases are compressed to the point that the separation dips below the equilibrium separation of the Morse bond. 
Physically, this corresponds to forcing the electron clouds of the atoms in the terminal bases to overlap.
At large separations, the DNA strand is stretched nearly straight and the harmonic bonds between adjacent bases expand past their equilibrium distance, leading to large mean forces for $x\geq132$.
This corresponds to stretching the bonds in the backbone of the linear DNA strand.

With these observations in mind, we chose the following fitting form for the mean force function,
\begin{align}
\bar{F}(x,T)=&-2 \alpha'  D' e^{\alpha'  (d_0'-x)}\left(e^{\alpha'  ({d_0'}-x)}-1\right)\nonumber\\
   &+2{K_s'} \left(x-R_0'\right)\cdot \Theta(x-R_0') \label{Ffit}\\
   &+T (c_1+c_2 x)\nonumber
\end{align}
Here, the first term is inspired by the Morse force $\vec{F}_M=-\nabla V_M$.
The primes on each variable denote that these are fitting parameters for the force function, not the parameters from the original potentials. 
The second term arises from the harmonic force $\vec{F}_H=-\nabla V_H$, which becomes prominent when the strand is stretched beyond the straight strand equilibrium length, defined as $R_0=(N-1)r_0$. 
This terms contains a Heaviside theta function $ \Theta(x-R_0')$ that ensures only the right half of the harmonic potential plays a role.
The fitted function requires this factor because intermediate separation values do not necessarily compress the harmonic bonds, since the chain may assume twisted conformations in three dimensional space.
The final term in the right hand side of equation~\ref{Ffit} generates the roughly linear temperature dependence shown in Figs.~\ref{fig:FbarVsTemp10-152} and~\ref{fig:FbarVsTempAll}. 
Since the slope of $\bar{F}$ versus $T$ is a strongly dependent and seemingly monotonically increasing function of $x$, we model it as a simple linear function with an intercept $c_1$ and a slope $c_2$.

Using Mathematica's \verb+FindFit+ function, we determined best fit values for the fitting parameters, which can be found in Table~\ref{Table1}.

\begin{figure}[b]
\includegraphics[scale=0.35]{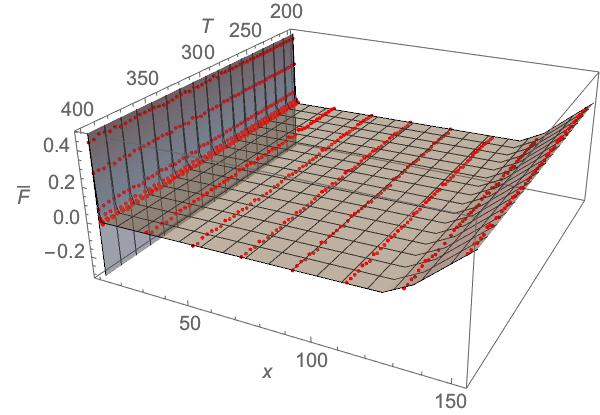}
\caption{\label{fig:FbarEqnOfState} A three dimensional plot of mean force $\bar{F}$ as a function terminal base pair separation $x$ and temperature $T$ (red) along with the corresponding fitted function (gray).}
\end{figure}

In Fig.~\ref{fig:FbarEqnOfState}, we make a three-dimensional plot of the mean force $\bar{F}$ versus separation $x$ and temperature $T$.
This plot shows the discrete simulation data (red) and the continuous fitted function (gray), which visually matches the data fairly well.

\begin{center}
 \begin{table}
 \begin{tabular}{|c |  c|} 
 \hline
 Parameter & Fit Value \\ [0.5ex] 
 \hline\hline
$\alpha'$ & 4.43  \\
\hline
$D'$ & 0.220 \\
\hline
$d_0'$ & 10.0  \\
\hline
$K_s'$ & 0.00603  \\
\hline
$R_0'$ & 129 \\
\hline
$c_1$ & $-1.42\times 10^{-5}$   \\
\hline
$c_2$ & $  4.74\times 10^{-7}$  \\ [1ex] 
 \hline
\end{tabular}
\caption{\label{Table1} Fitting parameters for the $\bar{F}(x,T)$ function.}
\end{table}
\end{center}

\begin{figure}[h]
\includegraphics[scale=0.35]{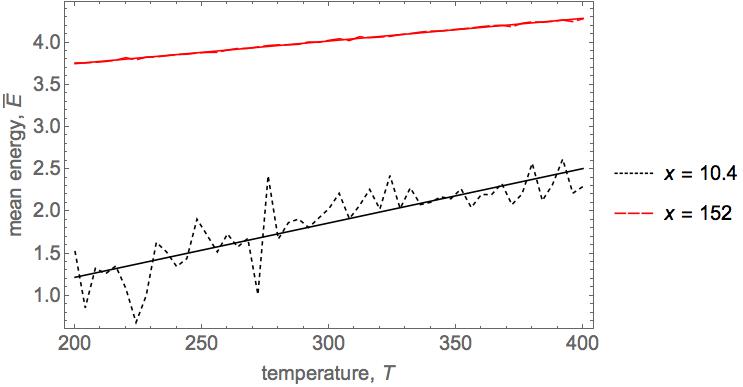}
\caption{\label{fig:EbarVsT10-152} A plot of the mean energy $\bar{E}$ as a function of temperature $T$ for terminal base pair separation values of $x = 10.4$ and 152 as temperature increases.}
\end{figure}

As with the mean force on the terminal base, it is beneficial to fit a functional form to the mean energy of the system.
In Fig.~\ref{fig:EbarVsT10-152}, we plot the mean energy $\bar{E}$ as a function of temperature $T$ for terminal base pair separation values $x = 10.4$ and 152.
As with the mean force function, there is a significant difference in slopes, and additionally, a stark contrast in the noise at these two separation values.
Again, observing plots of $\bar{E}$ versus $T$ for many separation values helps elucidate trends in the data.

\begin{figure}[b]
\includegraphics[scale=0.35]{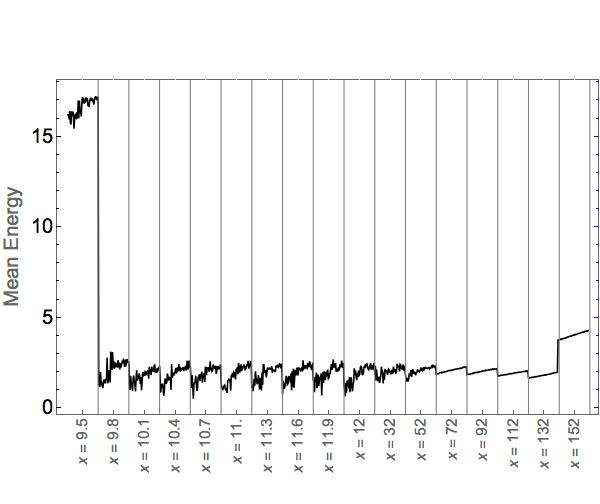}
\caption{\label{fig:EbarVsTAll} A plot of the mean energy $\bar{E}$ as a function of temperature $T$ for terminal base pair separations ranging from $x = 9.5$ to $x=152$. Note that the slope for all values of $x$ are positive.}
\end{figure}

\begin{figure}[t]
\includegraphics[scale=0.35]{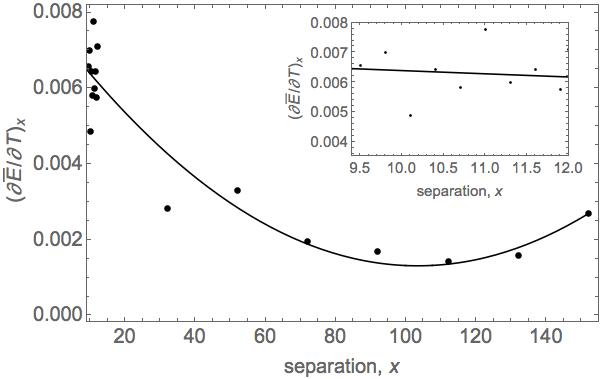}
\caption{\label{fig:SlopeFit} A plot of the derivative of the energy with respect to the temperature $\left(\partial\bar{E}/\partial T\right)_x$ versus separation $x$. The solid lines represents a quadratic fit to the data. The inset shows the same data near the inside of the Morse well from $x=9.5$ to $x=12$.}
\end{figure}

In Fig.~\ref{fig:EbarVsTAll}, we plot the mean energy $\bar{E}$ as a function of $T$ for multiple separation values.
In contrast to Fig.~\ref{fig:FbarVsTempAll}, the slope is positive for all $x$ values, which indicates that energy increases as temperature rises as expected. 
Furthermore, the slope of these plots does not appear to monotonically increase as a function of $x$, which is made clearer by the plot of fitted slope $\left(\partial\bar{E}/\partial T\right)_x$ versus $x$ shown in Fig~\ref{fig:SlopeFit}.
The energy at the extreme values of $x$ is noticeably larger due to the compressing of the Morse bond and stretching of harmonic springs.
The compressed states at $x = 9.5$ and $x=9.8$ may contain a slight s-shape reminiscent of a melting curve
This suggests there may be a pseudo-phase transition between the hybridized and unhybridized states as one increases the temperature, however it is difficult to say for certain because of the large amount of noise. 
Slightly larger values of $x$ outside the Morse well appear mostly linear, though there may be some downward curvature masked by the noise. 
Highly stretched hairpins (e.g. $x\geq72$) exhibit a lower noise straight line dependence on the temperature. 
This can be easily explained; at this extreme, the energy is dominated by the energy of the harmonic bonds, which have a linear dependence on the temperature according to the equipartition theorem. 
Stretching the strand length past the straight strand equilibrium length $R_0$ results in a large harmonic bond energy, as can be seen in from the data at $x=152$.

We fit the mean energy $\bar{E}$ to the function, 
\begin{align}
\bar{E}(x,T)=&{D''} \left(e^{\alpha''  ({d_0''}-x)}-1\right)^2\nonumber\\
		   &+ {K_s''} (x-R_0'')^2\cdot\Theta (x-R_0'')\label{eqn:Efit}\\
		   &+Tm(x)+b(x) \nonumber
\end{align}
Here, the double primes on the parameters indicate that they are mean energy fitting parameters, not the parameters defined for the original model.
The first term on the right hand side mimics the energy due to the Morse potential, and the second term mimics the energy due to the harmonic potential.
As with our mean force function, we included a factor containing a Heaviside theta function to truncate the left side of the harmonic well, which does not play a significant role.
The final two terms on the third line of equation~\ref{eqn:Efit} provide the temperature dependence.
Given the large amount of noise at small separations, we chose to err on the side of simplicity and use a linear fit to the temperature, albeit one whose slope $m(x)=m_0+m_1x+m_2x^2$ and intercept $b(x)=b_0+b_1x+b_2x^2$ are fitted to quadratic functions of the separation $x$.

\begin{center}
 \begin{table}[H]
  \begin{tabular}{||c |c |  |c| c||} 
 \hline
 Parameter & Fit Value  & Parameter & Fit Value \\ [0.5ex] 
 \hline\hline
$\alpha''$ & 5.19 & $m_0$ & $7.53\times 10^{-3}$ \\
\hline
$D''$ & $8.98\times 10^{-2}$  & $m_1$ & $-1.21\times 10^{-4}$  \\
\hline
$d_0''$ & 10.0 & $m_2$ & $5.83\times 10^{-7}$   \\
\hline
$K_s''$ & $ 1.23\times 10^{-2}$  & $b_0$ & $  -0.554$  \\  
\hline
 $R_0$ & $  138$  & $b_1$ & $  4.45\times 10^{-2}$  \\ 
 \hline
 & & $b_2$ & $  -2.38\times 10^{-4}$ \\ [1ex] 
 \hline
\end{tabular}
\caption{\label{Table2} Fitting parameters for the $\bar{E}(x,T)$ function.}
\end{table}
\end{center}

\begin{figure}[t]
\includegraphics[scale=0.35]{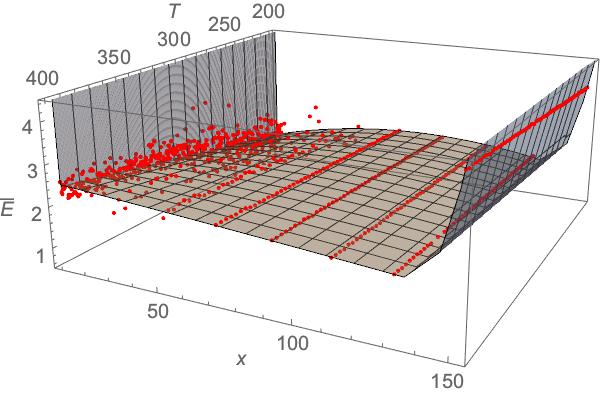}
\caption{\label{fig:EbarEqnOfState} A three dimensional plot of mean energy $\bar{E}$ as a function terminal base pair separation $x$ and temperature $T$ (red) along with the corresponding fitted function (gray).}
\end{figure}

To determine the fitting parameters, we again used Mathematica's \verb+FindFit+ function.
The fitting results can be found in Table~\ref{Table2}.
In Fig.~\ref{fig:EbarEqnOfState}, we plot the discrete simulation data (red) and smooth fitted function (gray), which visually appears to match the simulation data within some uncertainty.

After obtaining these state equations, it is straightforward to obtain the thermodynamic response functions.
In the following sections, we compute the isothermal compressibility, thermal expansion coefficient, heat capacity at constant separation, and the shapes of the adiabatic and isothermal pathways.

\subsection{Isothermal Compressibility, $\kappa_T$}%%%%%%%%%%%%%%%%%%%%%%%%%%%%%%%%%%%%%

After obtaining the equations of state for the DNA hairpin, it is straightforward to compute the response functions.
Here, we define the isothermal compressibility for a single DNA hairpin as 
\begin{equation}
\kappa_T= \left(\frac{\partial x}{\partial \bar{F}}\right)_T
\label{eqn:kappa}
\end{equation}
Note that our definition does not include the inverse length $1/x$ factor, which is conventional when discussing the compressibility of bulk materials.

\begin{figure}[h]
\includegraphics[scale=0.35]{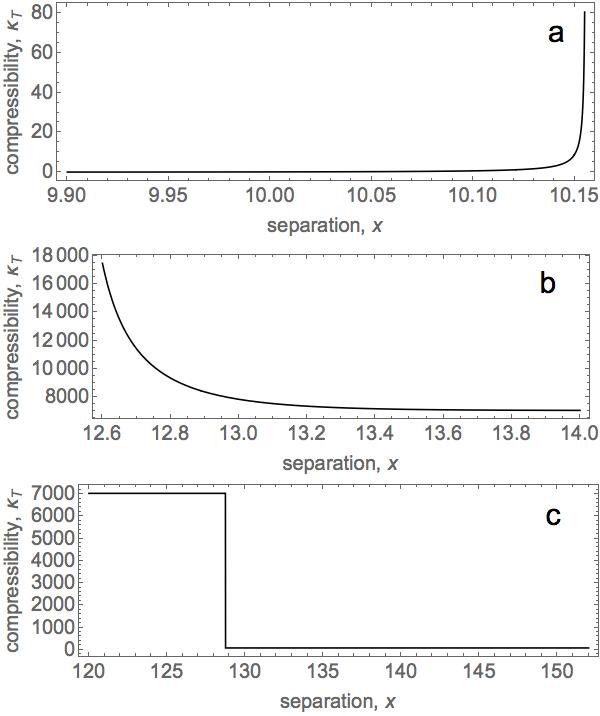}
\caption{\label{fig:CompressibilityAll} A plot of the isothermal compressibility $\kappa_T$ versus terminal base pair separation $x$ for (a) $x\le 10.2$, (b) $12.6 \le x \le 14$ and (c) $x \ge120$. }
\end{figure}

We computed the isothermal compressibility by taking the inverse of the partial derivative of the fitted function for the mean force $\bar{F}(x,T)$ with respect to $x$.
In Fig.~\ref{fig:CompressibilityAll}, we plot the compressibility as a function of terminal base pair separation for a variety of separation ranges.
In the region $x\lesssim10$, one can observe a small compressibility on the order $10^{-2}$ to $10^{-1}$.
In this range, the Morse bond of the terminal base pair is already somewhat compressed.
Experimentally, further reduction in the bond length leads to partial overlap of the electron clouds in atoms of the terminal bases, which requires a large force and results in a small compressibility.

The compressibility changes dramatically as the terminal bases are stretched to a separation $x\gtrsim10.15$. 
In this region, the hairpin becomes unstable, and the compressibility approaches infinity as shown in Fig.~\ref{fig:CompressibilityAll}A.
This limit represents the cusp of the bond between the terminal bases breaking.
When stretched past this separation, the hairpin pops open in a discontinuous transition to the open state.
For separations $x\approx 12.6$, the compressibility again approaches infinity as shown in Fig.~\ref{fig:CompressibilityAll}B.
This point represents the limit at which the open DNA strand snaps shut into its hairpin form.
The unstable region, in which the compressibility becomes negative, extends over a range $10.2 \lesssim x \lesssim12.5$.

\begin{figure}[b]
\includegraphics[scale=0.35]{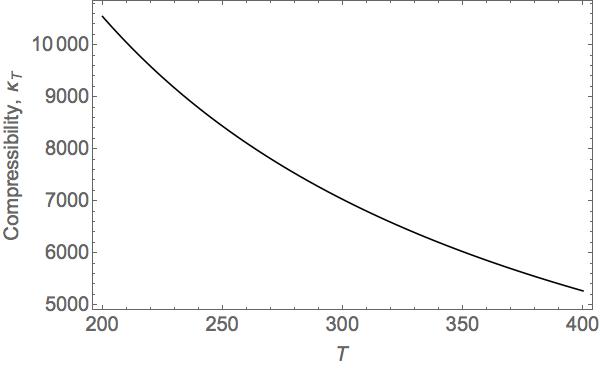}
\caption{\label{fig:Compressibility4} A plot of the isothermal compressibility $\kappa_T$ versus temperature $T$ for $\bar{F}=0$.}
\end{figure}

In the separation range $14<x<120$, the DNA strand is open but not stretched to its full extent.
Over this range, the compressibility does not have a significant dependence on the separation value $x$.
The compressibility varies with temperature over a range $5000<\kappa_T<10,000$ for temperatures in the range $200<T<400$, as can be seen in the plot of Fig.~\ref{fig:Compressibility4}.

The compressibility is plotted for separations $x\gtrsim120$ in Fig.~\ref{fig:CompressibilityAll}C.
Here, the DNA strand is nearly stretched to its full extent.
The harmonic bonds connecting adjacent bases in the strand stretch as the terminal ends are pulled, forcing the strand to an almost straight configuration.
In this arrangement, it takes an exceedingly large force to stretch the DNA any further from its already highly strained configuration.
For this reason, the compressibility drops rapidly to a roughly constant value around $\kappa_T\approx 80$.

\subsection{Thermal Expansion Coefficient, $\alpha$}%%%%%%%%%%%%%%%%%%%%%%%%%%%%%%%%%%%%%

We define the thermal expansion coefficient as follows,
\begin{equation}
\alpha= \left(\frac{\partial x}{\partial \bar{T}}\right)_{\bar{F}}=-\left(\frac{\partial x}{\partial \bar{F}}\right)_T\left(\frac{\partial \bar{F}}{\partial T}\right)_x.
\label{eqn:alpha}
\end{equation}
As with the isothermal compressibility, we neglect the factor of $1/x$ which is included in the standard definition of the thermal expansion coefficient for bulk materials.
The thermal expansion coefficient can be computed directly by taking partial derivatives of the fitted force function $\bar{F}(x,T)$ from equation~\ref{Ffit} with respect to $x$ and $T$ and using the cyclic rule, as shown in the right hand side of equation~\ref{eqn:alpha}.

\begin{figure}[h]
\includegraphics[scale=0.35]{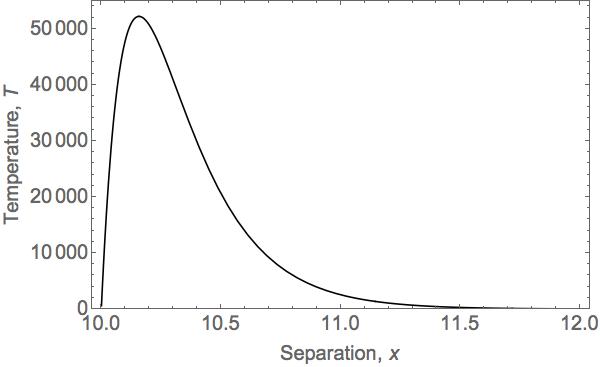}
\caption{\label{fig:ThermExpCoeff1} A plot of temperature $T$ versus separation $x$ for a force fixed at $\bar{F}=0$.}
\end{figure}

\begin{figure}[h]
\includegraphics[scale=0.35]{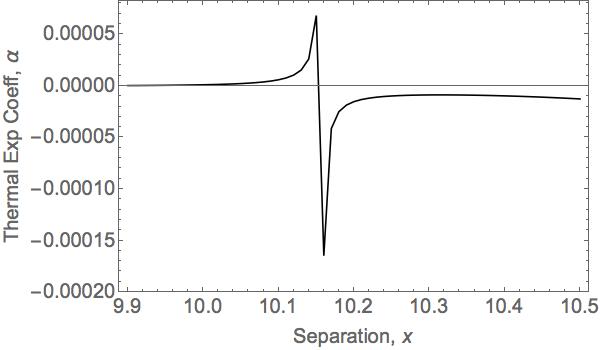}
\caption{\label{fig:ThermExpCoeff4} A plot of thermal expansion coefficient $\alpha$ versus separation $x$ for a force fixed at $\bar{F}=0$.}
\end{figure}

To help interpret the thermal expansion data, we first plot the temperature $T$ as a function of the terminal base pair separation $x$ for a fixed force $\bar{F}=0$ in Fig.~\ref{fig:ThermExpCoeff1}.
From the figure, we see that the slope of the graph is positive for separations smaller than $x\approx10.15$, which produces a positive thermal expansion coefficient.
At a separation $x\approx10.15$, there exists a peak giving a slope of zero, and a thermal expansion coefficient that approaches infinity as illustrated in Fig.~\ref{fig:ThermExpCoeff4}.
For separations in the range $x\gtrsim 10.15$, the slope is negative, giving a negative thermal expansion coefficient (Fig.~\ref{fig:ThermExpCoeff4}).
As with the compressibility, this region is unstable.
Simply put, under zero applied force, the hairpin can only thermally expand so far before popping open.
This makes sense given the extremely large temperature values near the peak. 
The peak in Fig.~\ref{fig:ThermExpCoeff1} corresponds to a spinodal, which is the limit of metastability. 
This is somewhat expected, since spinodals have been shown to exist in mean field models of DNA~\cite{santos_kinetic_2013}.
Points to the immediate left of that peak are metastable with respect to the open state of the DNA strand.

\begin{figure}[h]
\includegraphics[scale=0.35]{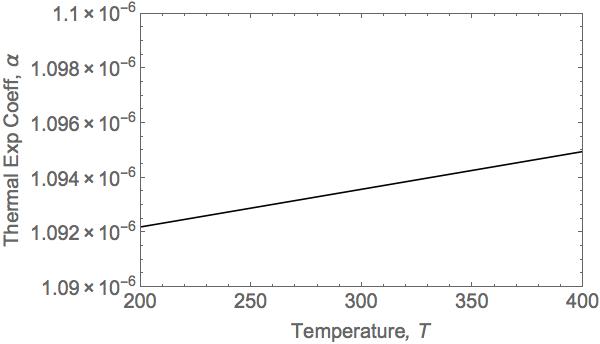}
\caption{\label{fig:ThermExpCoeff2} A plot of thermal expansion coefficient $\alpha$ versus temperature $T$ for a force fixed at $\bar{F}=0$.}
\end{figure}

In Fig.~\ref{fig:ThermExpCoeff2}, we plot the thermal expansion coefficient as a function of temperature in the range from $T=200$ to 400 for a fixed force $\bar{F}$ = 0. 
As can be seen from the plot, the thermal expansion coefficient varies little over this temperature range and remains on the order $\alpha\approx10^{-6}$.
This is in sharp contrast, to Fig.~\ref{fig:ThermExpCoeff4} in which $\alpha\rightarrow\infty$.
However, the temperature range over which the thermal expansion coefficient blows up is much larger than what is biologically relevant or even technologically feasible.
As such, we can assume that for reasonable temperature values, the model predicts a roughly constant thermal expansion coefficient.

\subsection{Constant Separation Heat Capacity, $C_x$}%%%%%%%%%%%%%%%%%%%%%%%%%%%%%%%%%%%%%

\begin{figure}[H]
\includegraphics[scale=0.35]{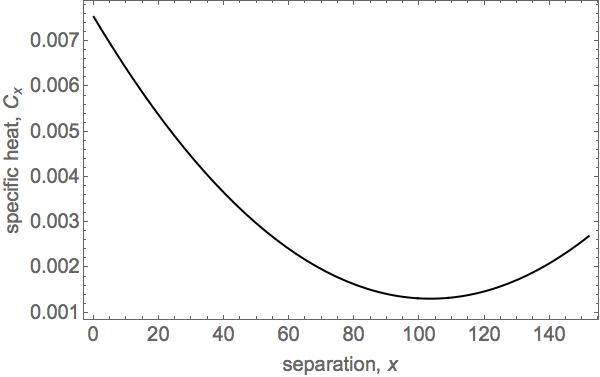}
\caption{\label{fig:SpecificHeatConstX} A plot of the specific heat $C_x$ versus separation $x$.}
\end{figure}

We define the constant separation heat capacity as,
\begin{equation}
C_x= \left(\frac{\dbar Q}{dT}\right)_x=\left(\frac{\partial \bar{E}}{\partial T}\right)_x,
\label{eqn:Cx}
\end{equation}
where $\dbar Q$ is the differential heat added to the system.
We computed the partial derivative of the mean energy with respect to temperature directly using the fitted form for $\bar{E}$ from equation~\ref{eqn:Efit}.
Since the fitted mean energy depends linearly on the temperature, there is no temperature dependence in the specific heat.
We plot the $C_x$ as a function of separation $x$ in Fig.~\ref{fig:SpecificHeatConstX}.

In Fig.~\ref{fig:SpecificHeatConstX}, we see that the specific heat at constant separation exhibits a minimum at intermediate values of stretching. 
For DNA in the hairpin configuration (i.e. small $x$), this makes sense, since unstretched strands require more thermal energy to break bonds. 
Similarly, highly stretched strands are strained and require a large amount of energy to further increase thermal fluctuations.

%\subsection{Constant Force Specific Heat, $C_F$}%%%%%%%%%%%%%%%%%%%%%%%%%%%%%%%%%%%%%

\subsection{Isothermal and Adiabatic Pathways}%%%%%%%%%%%%%%%%%%%%%%%%%%%%%%%%%%%%%

By definition,  adiabatic pathways feature no heat transfer, $\Delta Q = 0$.
From the first law of thermodynamics, we may write
\begin{align}
d\bar{E}= \dbar Q+\dbar W = \dbar Q+\bar{F} dx,
\label{eqn:FirstLaw}
\end{align}
where  $\dbar W=\bar{F} dx$ is the quasistatic differential work done on the DNA strand.
Along an adiabatic path, it must be true that $d\bar{E}= \bar{F} dx$.
To determine the shape of an adiabatic path for a DNA hairpin, we use the following procedure.
\begin{enumerate}
\item We first chose an initial temperature $T_i$ and terminal base pair separation $x_i$ as our starting point, and from these, we computed the mean force $\bar{F}_i$.
\item We then chose the next value for the mean separation,
\begin{equation}
x_{i+1}=x_i+dx,
\end{equation}
where we chose $dx=0.01$ small enough to give a smooth curve.
\item We found $T_{i+1}$, such that,
\begin{align}
\bar{E}(x_{i+1},T_{i+1})&-\bar{E}(x_{i},T_{i})=\nonumber\\
&\left(\frac{\bar{F}(x_{i+1},T_{i+1})+\bar{F}(x_{i},T_{i})}{2}\right)dx
\label{eqn:adiabat}
\end{align}
\item Finally, steps 1 through 3 were iterated many times to determine the shape of the adiabatic pathway.
\end{enumerate}
Equation~\ref{eqn:adiabat} is simply a discretized version of equation~\ref{eqn:FirstLaw}, in which we have averaged of the mean forces between points and set $\dbar Q$ to zero.

\begin{figure}[H]
\includegraphics[scale=0.35]{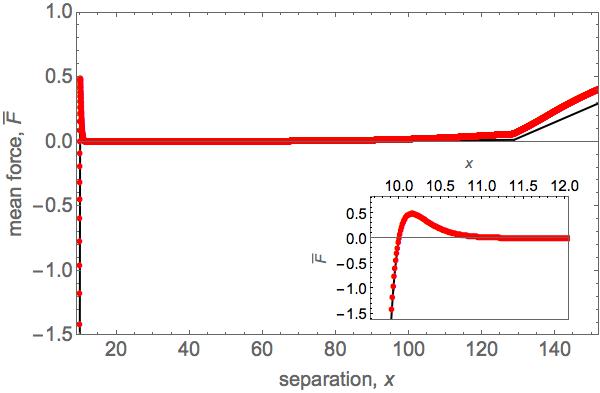}
\caption{\label{fig:Adiabatic1} A plot of the mean force $\bar{F}$ versus separation $x$ illustrating the adiabatic (red) and isothermal (black) pathways. The adiabatic and isothermal pathways only differ significantly when the DNA strand is greatly stretched.}
\end{figure}

In Fig.~\ref{fig:Adiabatic1}, we plot the isothermal (black, solid) and adiabatic (red, dotted) pathways.
As can be seen from the inset on the figure, inside the well within the region from $x<12$, there is little difference between the adiabatic and isothermal curves.
This is likely due to the large force associated with the Morse bond between the terminal base pair.
Only when the hairpin is stretched to its full linear extent $x\gtrsim130$ do we observe an appreciable deviation between the adiabatic and isothermal pathways.

%%%%%%%%%%%%%%%%%%%%%%%%%%%%%%%%%%%%%%%%%%%%%%%%%%%%%%%%%
\section{Conclusions}

In summary, we used the Metropolis Monte Carlo algorithm to simulate a three dimensional version of the CLPG model.
We obtain fitted equations of state for the mean force and energy of DNA hairpins. 
From these equations of state, we computed several response functions and determined the shape of the adiabatic pathway.
The equations of state and response functions are potentially useful thermodynamic properties of the hairpin. 
For this reason, the results presented here may be used to inform researchers studying biological processes or developing novel nanoscale devices.

Many reactions and interactions in biochemical processes take advantage of chemical energy. 
For example, proteins achieve motion by binding to ATP, which spontaneously dissociates into ADP and a phosphate ion, resulting in repulsive forces that convert chemical energy to kinetic energy. 
By imitating the mechanics of these mechanical biological processes, molecular nanomachines could be designed to achieve a desired motion. 
A DNA hairpin that opens and closes could be used to move molecular scale objects similar to the snapping-open of ATP. 
Unlike ATP, which utilizes chemical energy, the mechanics of the hairpin can be controlled by macroscopically by adjusting the temperature, making it more akin to a heat engine than a chemically-powered engine.

The creation of a nanoscale heat engine would not be new. 
Other forms of nanoscale heat engines include Otto engines \cite{abah_single-ion_2012, rossnagel_nanoscale_2014}, all-optical nanomechanical heat engines \cite{dechant_all-optical_2015}, and cold-atom based heat engines \cite{brantut_thermoelectric_2013}. 
%The Otto heat engine is a theoretical heat engine that uses a single ion and a Paul trap. 
%This type of heat engine requires an ion to be completely isolated from its environment; the heat bath is substituted by a reservoir of light controlled via Doppler cooling. 
%The all-optical nanomechanical heat engine involves levitating a nanoscale object using an optical trap that creates a harmonic potential in a moderate vacuum. The vacuum contains the optical trap and a heat bath composed of the rest gas inside the trap. 
%The cold-atom based heat engine generates a particle current between two atomic reservoirs connected by a mesoscopic channel. A temperature bias generates the particle current in the direction from the hot reservoir to the cold reservoir, this happens due to the thermoelectric properties of the channel.   
Unlike these examples, a DNA-based heat engine would be able to operate in ambient solution at or above room temperature, which may make it more accessible experimentally.

Piezochromic luminescent materials are now being used in some nanoscale heat engines to determine whether the heat engine has under gone the desired mechanical stimuli \cite{sagara_mechanically_2009, kunzelman_oligo_2008}. Piezochromic materials fluoresce when they undergo dynamic phenomenon such as shearing, grinding or pressure  \cite{kunzelman_oligo_2008}. 
By attaching piezochromic materials to the terminal ends of the DNA hairpin, one (a) should be able determine whether the hairpin opened or not by detecting the fluorescence and (2) could potentially extract useful energy from the light emitted.

%%%%%%%%%%%%%%%%%%%%%%%%%%%%%%%%%%%%%%%%%%%%%%%%%%%%%%%%%

%Chi2 fit test
%Discuss fitting parameters
%cite the important of spinodals in DNA

\begin{acknowledgments}
This work was done as part of Simpson College's PHY-310 Undergraduate Thermal Physics class. 
We would like to thank Simpson College President Jay Simmons, Dean Steve Griffith, and Physics Department Chair David Olsgaard for their support and the use of the Physics Department laptops. We would also like to thank Professor Derek Lyons for useful conversations. 
\end{acknowledgments}

%\nocite{*}

%\bibliographystyle{plain}
\bibliography{refthermal2}% Produces the bibliography via BibTeX.

\end{document}